\documentclass[%
 aip,
 amsmath,amssymb,
 reprint,%
]{revtex4-1}

\usepackage{graphicx}
\usepackage{dcolumn}
\usepackage{bm}
\usepackage{romannum}

\usepackage[T1]{fontenc}
\usepackage{mathptmx}
\begin{document}
\title[]{Pattern formation and chimera states in 2D SQUID metamaterials}
\author{J. Hizanidis}
\author{N. Lazarides}%
\author{G. P. Tsironis}
\affiliation{Department of Physics,  University of Crete, Herakleio, 71003, Greece\\
National University of Science and Technology MISiS, Leninsky Prospect 4, 
Moscow, 119049, Russia
}%
\email{hizanidis@physics.uoc.gr}
\date{\today}
\begin{abstract} 
The Superconducting QUantum Interference Device (SQUID) is a highly nonlinear 
oscillator with rich dynamical behavior, including chaos. When driven by a 
time-periodic magnetic flux, the SQUID exhibits extreme multistability at 
frequencies around the geometric resonance which is manifested by a 
``snake-like'' form of the resonance curve. Repeating motifs of SQUIDs form 
metamaterials, i.~e. artificially structured media of weakly coupled discrete 
elements that exhibit extraordinary properties, e.~g. negative diamagnetic 
permeability. We report on the emergent collective dynamics in two-dimensional 
lattices of coupled SQUID oscillators, which involves a rich menagerie of 
spatio-temporal dynamics, including Turing-like patterns and chimera states. Using Fourier analysis we characterize these patterns and identify characteristic spatial and temporal 
periods. In the low coupling limit, the Turing-like patterns occur near the synchronization-desynchronization 
transition which can be related to the bifurcation scenarios of the single SQUID. 
Chimeras emerge due to the multistability near the geometric resonance, and by varying the dc component of the external force we can make them appear and reappear and, 
also, control their location. 
A detailed analysis of the parameter space reveals the coexistence of Turing-like patterns and chimera states in our model, as well as the ability to transform between these states by varying the system parameters.

\end{abstract}
\maketitle
{\bf 
The Superconducting QUantum Interference Device (SQUID), hereafter referred to 
as ``SQUID'', is a highly nonlinear oscillator that exhibits strong resonant 
response to external magnetic fields. Its dynamics shows a wealth of phenomena
such as hysteresis, multistability, subharmonic resonances, saddle-node and 
period-doubling bifurcations and chaos, which can be revealed through its complex 
bifurcation structure. Its resonance curve, in particular, acquires a ``snake-like'' 
shape around its geometric resonance.

When many SQUIDs are arranged on a periodic array, they form magnetic metamaterials 
with extraordinary electromagnetic properties such as negative permeability, 
broad-band tunability, self-induced broad-band transparency, dynamic multistability 
and switching, as well as coherent oscillations. Besides their appeal as 
superconducting devices, SQUID metamaterials provide a unique testbed for exploring 
complex spatiotemporal dynamics. Here we demonstrate numerically that two-dimensional 
SQUID metamaterials (SQUID metasurfaces) support the emergence of certain spatially 
non-homogeneous dynamic states such as chimera states and patterned states of the Turing type.

Chimera states in SQUID metasurfaces make themselves apparent as domains of
SQUIDs with synchronized (coherent) dynamics that coexist with domains of
SQUIDs with desynchronized (incoherent) dynamics. Our system is an excellent
physical and technologically relevant example of a 
driven system, where studies on chimeras are limited.
Turing patterns, on the other
hand, typically hexagons, stripes, rhombi, or labyrinths, emerge in reaction
-diffusion systems. Since the SQUIDs in a metasurface are diffusively coupled,
the emergence of Turing-like patterns are shown, for the first time, to be possible. Our investigation relies on a well-established model for SQUID 
metasurfaces, whose parameters lie in the experimentally accessible ranges of 
the applied constant (dc) flux as well as the amplitude and the frequency of the 
applied time-dependent (ac) flux. We present numerically generated chimera 
states using appropriately selected initial conditions, as well as several 
Turing-like patterns whose characteristic length is determined from their 
corresponding two-dimensional spatial Fourier transforms. Moreover, in the low coupling limit, the region 
of stability for Turing-like patterns is related to the saddle-node bifurcation lines 
of the reduced equations for the SQUID metasurface.      

The interplay between chimera states and Turing-like patterns for such a system of driven nonlinear oscillators is being addressed for the first time. Through a detailed analysis of the parameter space we reveal the coexistence and transformation between these states in certain regions. 
}
\section{Introduction}
\label{sec1}
Superconducting metamaterials comprising Superconducting QUantum Interference 
Devices (SQUIDs), are artificial materials that exhibit exceptional properties 
not found in nature, such as negative magnetic permeability, dynamic 
multistability \cite{Jung2014a,Zhang2015}, broadband tunability, and self-induced 
broadband transparency \cite{Zhang2015}. Some of these extraordinary properties 
have been predicted theoretically both for the quantum \cite{Du2006} and the 
classical regime \cite{Lazarides2007,Lazarides2013,Lazarides2018}. They can be 
implemented and studied in various designs and arrangements, both in one and two 
dimensions \cite{Trepanier2013,Zhang2015,Zhang2016,Butz2013,Jung2014a,Jung2014b,
Ustinov2015,KIS19}. Recently, the degree of spatio-temporal coherence of SQUID
metamaterials was examined experimentally and numerically using microwave 
transmission measurements~\cite{Trepanier2017}. Moreover, its quantum counterpart, 
the qubit, has been proposed as an important ``building block'' of quantum 
computers~\cite{SAI13,SHU18}.

Apart from their technological applications, SQUID metamaterials provide a unique 
testbed for exploring complex spatio-temporal dynamics. A SQUID metamaterial is 
in essence a system of nonlinear coupled oscillators with inertia, which are 
driven, damped, and are characterized by a nonlinear term which enters through 
the Josephson effect \cite{Josephson1962}. A very prominent dynamical feature of 
SQUID metamaterials are dissipative breathers~\cite{LAZ08}, which emerge as a 
result of their discreteness, the weak coupling between their elements and the 
Josephson nonlinearity.

In this work, we will study the collective behavior beyond these localized 
states, in a dynamical regime where the whole lattice is oscillating and is 
capable of creating Turing-like patterns \cite{Turing1952}. Pattern forming 
systems have been in the center of scientific research for decades in a rich 
variety of natural and laboratory scenarios~\cite{CRO93}. These include 
oscillatory chemical reactions governed by reaction-diffusion dynamics~
\cite{SHO15}, static and excitable biological media~\cite{KOC94}, dissipative 
spatio-temporal solitons in nonlinear optics~\cite{ARE99}, and many more. Here, 
we will use synchronization measurements in order to explore pattern formation 
in two-dimensional (2D) SQUID metamaterials.

Another phenomenon related to synchronization which has been observed in SQUID 
metamaterials are chimera states, where domains of coherent and incoherent 
motion coexist in an otherwise symmetric network of identical oscillators~
\cite{KUR02a,ABR04,panaggio:2015,YAO16,OME18}. SQUID chimeras have mainly been 
studied 
in one-dimensional (1D) arrays~\cite{LAZ15,HIZ16a,HIZ16b,BAN18b}. Here we will 
explore this phenomenon for locally coupled SQUIDs on a tetragonal lattice. 
Higher-dimensional chimeras have been the subject of recent works involving 
networks of Kuramoto and neuronal oscillators~\cite{MAI15,SCH17,KAS18,ARG19}. 
The interplay, however, between chimeras and Turing-like patterns has not been 
addressed sufficiently and this a new element that our work focuses on. 
Note that the system under study is a physical,
technologically relevant example of a \emph{forced} system, where such dynamics is still to be investigated ~\cite{DUD16,CLE18}. 

\section{Single SQUID dynamics}
\begin{figure}[ht]
\includegraphics[width=\columnwidth]{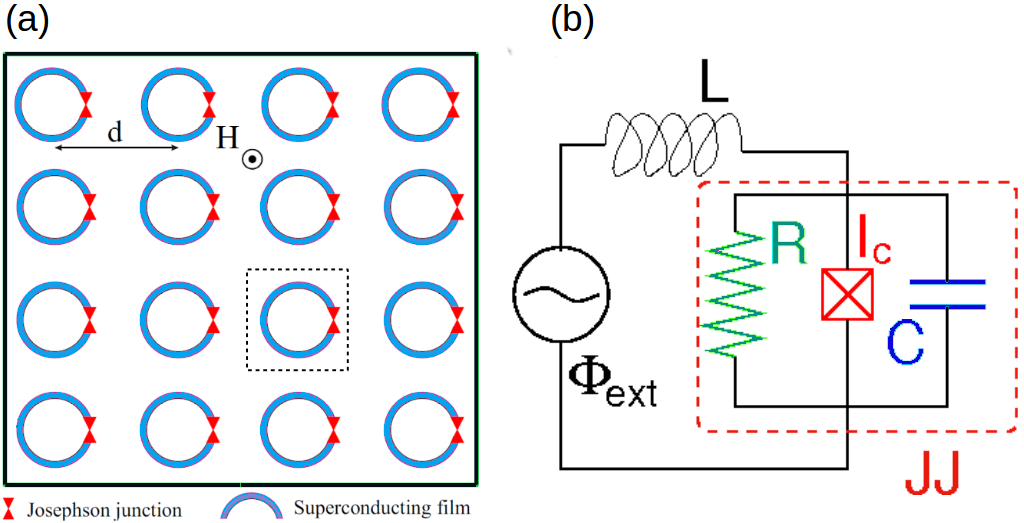}
\caption{
(a) Schematic of a two-dimensional SQUID metamaterial in a magnetic field $H(t)$ 
    and 
(b) equivalent electrical circuit of the single SQUID (marked by the dashed box 
    in (a)) in the RCSJ framework.}
\label{fig1}
\end{figure}

A SQUID consists of a superconducting ring interrupted by a Josephson junction 
(JJ) as shown schematically inside the dashed box of Fig.~\ref{fig1}(a). When 
placed in a perpendicular, spatially uniform magnetic field $H$, a current $I$ 
is induced which is the sum of the supercurrent $I_s$ flowing through the JJ and
the quasiparticle current. Then, the magnetic flux $\Phi$ threading the loop of 
the SQUID is given by:
\begin{eqnarray}
\label{eq1}
  \Phi =\Phi_{ext} +L\, I ,
\end{eqnarray}
where $L$ is the self-inductance of the SQUID ring and 
$\Phi_{ext} =\Phi_{dc} +\Phi_{ac} \, \cos( \omega t )$
is the external flux applied to the SQUID, containing both a constant (dc) flux 
bias $\Phi_{dc}$ and an alternating (ac) flux of amplitude $\Phi_{ac}$ and 
frequency $\omega$. 
\begin{figure*}[]
\includegraphics[width=\textwidth]{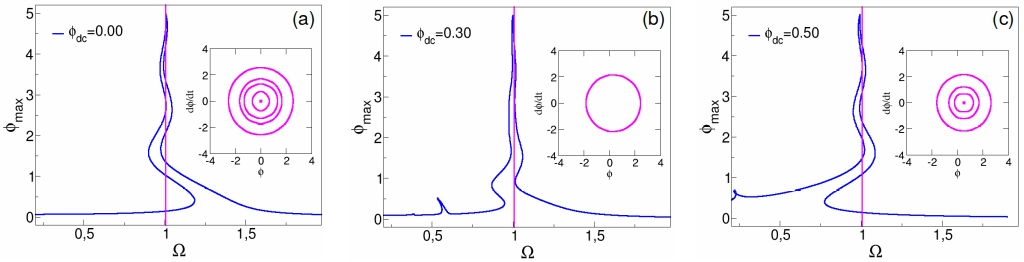}
\caption{Resonance curves of a single SQUID for various dc flux values:
(a) $\phi_{dc}=0.0$, (b) $\phi_{dc}=0.3$, and (c) $\phi_{dc}=0.5$.
The vertical line marks the value of the geometric resonance frequency 
and the insets show the corresponding stable periodic solutions at that value. 
Other parameters are: $\phi_{ac}=0.06$, $\gamma=0.024$ and $\beta=0.1369$.}
\label{fig02}
\end{figure*}

The current $I$ in the SQUID is given by the resistively and capacitively 
shunted junction (RCSJ) model of the JJ \cite{Likharev1986}, schematically shown 
in Fig.~\ref{fig1}(b):
\begin{eqnarray}
\label{eq2}
  I =-C\frac{d^2\Phi}{dt^2} -\frac{1}{R} \frac{d\Phi}{dt} 
     -I_c\, \sin\left(2\pi\frac{\Phi}{\Phi_0}\right), 
\end{eqnarray}
where $C$ is the capacitance of the JJ of the SQUID, $R$ is the resistance, 
$I_c$ is the critical current which characterizes the JJ, $\Phi_0$ is the flux 
quantum, and $t$ is the temporal variable. Combining Eqs. (\ref{eq1}) and 
(\ref{eq2}) we get:
\begin{eqnarray}
\label{eq3}
   \ddot{\phi} +\gamma \dot{\phi} +\phi +\beta \sin\left( 2\pi \phi \right) =
\phi_{dc}+ \phi_{ac} \cos(\Omega \tau), 
\end{eqnarray}
where all fluxes have been normalized to the flux quantum, $\phi=\Phi/\Phi_0$,
$\phi_{ac,dc}=\Phi_{ac,dc}/\Phi_0$, while the frequency and time variable have 
been normalized to the inductive-capacitive SQUID frequency,
$\omega_{LC}=1 / \sqrt{L C}$ and its inverse, respectively, i.~e. 
$\Omega={\omega}/{\omega_{LC}}$ and $\tau=t/\omega_{LC}^{-1}$. The parameter
$\beta=LI_c/\Phi_0 =\beta_L/2\pi$ is the so-called rescaled SQUID parameter
and $\gamma=\omega_{LC} L/R$ corresponds to the loss coefficient.

Typical values of the design parameters of a 
SQUID~\cite{Trepanier2013,Zhang2015} 
provide the dimensionless coefficients 
$\beta \simeq 0.1369$ ($\beta_L \simeq 0.86$) and $\gamma \simeq 0.024$ which 
appear in the normalized Eq. (\ref{eq3}) for the flux $\phi =\Phi / \Phi_0$ 
through the loop of the SQUID. They also provide experimentally plausible values
$f_{LC} =\omega_{LC} /(2\pi) \simeq 13.9 ~GHz$ ($\Omega \simeq 1$) and 
$f_{SQ} =\omega_{SQ} /(2\pi) \simeq 18.9 ~GHz$ 
($\Omega =\Omega_{SQ} \simeq 1.364$) for the geometric and the linear resonance 
frequency of the SQUID, respectively~
\cite{Butz2013,Trepanier2013,Zhang2015}. The values of the externally controlled 
parameters $\phi_{dc}$, $\phi_{ac}$, and $\Omega$ used here, are within the range 
of the experimentally accessible values, i.~e., $\phi_{dc}$ in the interval 
$[-1, 2]$ \cite{Trepanier2013}, $\phi_{ac}$ in the interval $[0.001, 0.18]$ 
\cite{Zhang2015}, and $\Omega$ in the interval 
$\frac{2\pi}{\omega_{LC}} [10, 22.5] ~GHz$ \cite{Trepanier2013}.

By expanding the sine nonlinearity in Eq.~\ref{eq3} in a Taylor series and 
keeping the cubic term only, the SQUID model reduces to the famous driven Duffing 
oscillator. The latter is known to exhibit a nonlinear frequency response, 
bistability, hysteresis phenomena, and chaotic behavior. Similarly, the SQUID is 
capable of demonstrating complex dynamics, but with additional features owing to 
its higher-order nonlinear term. For a certain range of parameters the SQUID 
exhibits a ``snake-like'' resonance curve in which multiple stable and unstable 
periodic orbits coexist and vanish through saddle-node bifurcations of limit 
cycles~\cite{HIZ16a,HIZ18,DIM82}. The detailed bifurcation structure for zero 
and finite dc flux was first reported in ~\cite{HIZ16a,HIZ18}. Here we significantly further 
this analysis and explore the stability of solutions in the \emph{full} 
$(\phi_{dc},\Omega)$ parameter space.
\begin{figure*}
\includegraphics[width=\textwidth]{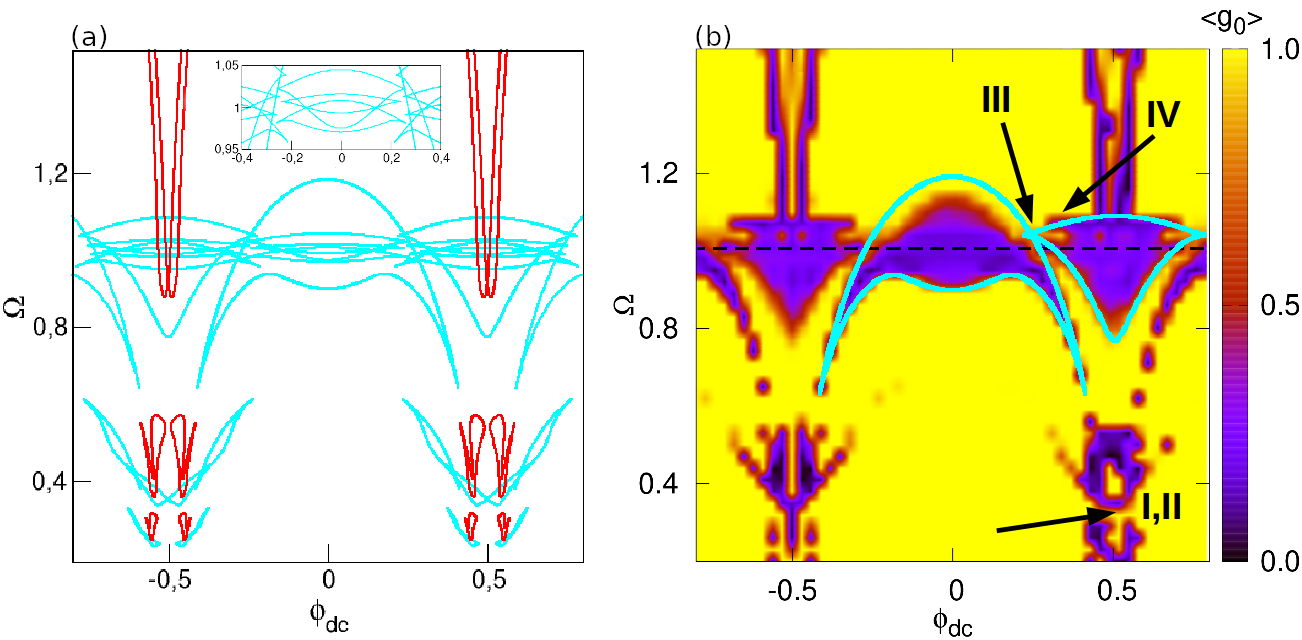}
\caption{
Left: Bifurcation diagram in the $(\phi_{dc},\Omega)$ plane of the single SQUID 
      oscillator. Blue and red lines correspond to saddle-node bifurcations of 
      limit cycles and period doubling bifurcations, respectively. The inset shows a blowup around $\Omega=1$.
Right: Value of the synchronization measure $\langle g_0 \rangle$ in the 
       $(\phi_{dc},\Omega)$ parameter space for a $30 \times 30$ SQUID lattice 
       with coupling strength $\lambda=-0.01$. The black and cyan curves 
       correspond to saddle-node bifurcation lines of the reduced system for 
       $\lambda=0.0$ and $\lambda=-0.01$, respectively.
Other parameters are: $\phi_{ac}=0.06$, $\gamma=0.024$ and $\beta=0.1369$.
\label{fig03}
}
\end{figure*}

Figure~\ref{fig02} shows the resonance curves of the single SQUID as the dc flux 
increases from 0 to 0.5. A saddle node 
bifurcation of limit cycles occurs at each turning point of the curve
where stable and unstable branches merge 
~\cite{HIZ16a,HIZ18}. The vertical line marks the geometric resonance 
frequency and the insets show the phase portraits of the corresponding \emph{stable} periodic solutions at that particular value of $\Omega$. 
As the dc flux increases, the ``center'' 
of these solutions shifts to the right and the number of coexisting limit cycles at $\Omega_{LC}$ changes.
For $\phi_{dc}=0.0$ (Fig.~\ref{fig02}(a)) we have five coexisting periodic solutions of different 
amplitudes, centered around the origin. As the dc flux increases, the number of coexisting orbits gradually shrinks to one, at $\phi_{dc}=0.30$, while new subresonances~\cite{MAR18,HIZ18} make an appearance at lower 
frequencies (Fig.~\ref{fig02}(b)). 
At $\phi_{dc}=0.50$ the SQUID is again 
multistable with four coexisting periodic solutions (Fig.~\ref{fig02}(c)) and, at the same time, the occurring saddle-node bifurcations have transformed from subcritical to supercritical and vice versa~\cite{HIZ18}.
This scenario repeats itself periodically with the resonance curve moving back and forth, with respect to $\Omega_{LC}$, as we vary $\phi_{dc}$.
This ``rocking'' of the snake-like 
resonance curve and, consequently, periodic appearance and disappearance of solutions around the geometric resonance frequency, is better visualized in the video SM1 of the Supplementary Material. 

All of the aforementioned features are reflected in 
Fig.~\ref{fig03}(a), where the co-dimension 2 bifurcation diagram in the 
$(\phi_{dc},\Omega)$ plane is depicted. The bifurcation lines have been obtained 
using a very powerful software tool that executes a root-finding algorithm for 
continuation of steady state solutions and bifurcation problems~\cite{ENG02}. 
Cyan and red lines denote saddle-node bifurcations of limit cycles and 
period-doubling bifurcations, respectively. The bifurcation structure is 
extremely delicate and periodic in $\phi_{dc}$ with a period of unity. This 
periodicity can be proven as follows: Assuming that $\phi$ is a solution of the 
single SQUID equation and by plugging $\phi \pm1$ into Eq.~\ref{eq3} we get: 
$d^2(\phi \pm 1)/dt^2 +\gamma d(\phi \pm 1)/dt +(\phi \pm 1) 
+\beta \sin\left( 2\pi \phi \pm 2\pi \right)=\phi_{dc}+\phi_{ac} \cos(\Omega \tau)$.
After simple manipulations we obtain: 
$\ddot{\phi}+\gamma \dot{\phi} +\phi +\beta \sin\left( 2\pi \phi \right) 
=(\phi_{dc}\pm 1)+\phi_{ac} \cos(\Omega \tau)$. Therefore, $\phi$ satisfies 
Eq.~\ref{eq3} also for a dc flux $\phi_{dc}\pm 1$, and $\phi_{dc}\pm 2$, and so on.

Looking at Fig.~\ref{fig03}(a) again, for fixed $\phi_{dc}$ values and moving in the 
$\Omega$ direction, we can recreate the resonance curves shown in Fig.~\ref{fig02} and the video SM1 of the Supplementary Material: 
The multiple and interwoven cyan lines correspond to the multiplicity of 
solutions around the geometric resonance frequency (better visible in the inset), while the red lines around 
$\phi_{dc}=0.5$ (and its symmetric $\phi_{dc}=-0.5$) are related to the 
subresonances that make their appearance for those dc flux values. The 
bifurcation diagram of Fig.~\ref{fig03}(a) presents additional, long period-doubling 
bifurcation branches extending to higher $\Omega$ values, which are not captured 
in the resonance curves of Fig.~\ref{fig02}. The period doubling lines are 
symmetrical around $\phi_{dc}=\pm0.5$ and for higher ac flux values are 
associated with corresponding chaotic regions, as shown in \cite{HIZ18}, where 
the maximum Lyapunov exponent was calculated in the $(\phi_{dc},\Omega)$ plane.


\section{Two-dimensional SQUID lattices}
In this work, we will focus on two regimes of the driving frequency: The vicinity 
of $\Omega_{LC}$, and at lower values around $\Omega=0.3$. Through the single 
SQUID complex dynamics, we aim at interpreting the collective behavior of the 2D
SQUID lattice. We consider a planar $N\times N$ SQUID array consisting of 
identical units as shown in Fig.~\ref{fig1}(a), arranged in an orthogonal lattice 
with a constant distance $d$ in both $x$ and $y$ directions. The induced current 
$I_{nm}$ produces a magnetic field which couples each SQUID with all the others 
due to magnetic dipole-dipole interactions through their mutual inductance. To a 
good approximation, we may assume that the SQUIDS are coupled only to their 
nearest neighbors, neglecting further-neighbor interactions. The dynamic 
equations for the normalized flux through the ring of the $(n, m)$-th SQUID, 
$\phi_{nm}$, are given by~\cite{LAZ08}:
\begin{eqnarray}
\label{eq04}
 &&\ddot{\phi}_{nm} +\gamma \dot{\phi}_{nm} 
   +\phi_{nm} +\beta \sin\left( 2\pi \phi_{nm} \right) 
   \nonumber \\
 &=&\lambda( \phi_{n-1,m} +\phi_{n+1,m}+\phi_{n,m-1} +\phi_{n,m-1} ) 
   \nonumber \\
 &+&(1 -4\lambda) (\phi_{dc}+\phi_{ac} \cos(\Omega \tau)), \quad n,m=1\dots N,
\end{eqnarray}
where $\lambda\equiv M /L$ is the coupling constant between any two neighboring 
SQUIDs, coupled through their mutual inductance $M$. The value of $M$ is negative 
due to the fact that the magnetic field generated by one SQUID crosses the 
neighboring SQUID in the opposite direction. In the following, we will study the 
nature of the synchronization-desynchronization transitions and will identify 
the collective states that emerge in relevant regimes of the parameter space.
The latter involves the two parameters which can be easily tuned in an 
experiment, namely the dc flux and the frequency of the ac flux, with the other 
parameters (ac flux amplitude $\phi_{ac}$, $\gamma$ and $\beta$) kept constant.

Equations~(\ref{eq04}) are integrated numerically in time using a standard 
fourth-order Runge-Kutta algorithm with a time-step equal to $0.02$ and periodic 
boundary conditions, i.~e., $\phi_n(\tau)=\phi_{N+n}(\tau)$ for all $n$. This particular choice of boundary conditions does not affect the dynamics significantly. By using instead, e.~g., free-end boundary conditions, only slight differences would have been observed which account for $1-2\%$ deviations of the corresponding $g_0$ value. Moreover, if nonlocal interaction between SQUIDs were assumed, a slight suppression of global synchronization would have been observed with both periodic and free-end boundary conditions.

The initial conditions for the 
$\phi_{nm}$ values follow a Gaussian random distribution in the interval $[-3,3]$ 
and $\dot \phi_{nm}=0.0$. We will 
employ a quantification measure, originally introduced for the classification 
for chimera states~ \cite{kevrekidis}, the \emph{local curvature}, which is 
calculated at each time instance by applying the absolute value of the discrete 
Laplacian on the spatial data of the magnetic flux:
\begin{eqnarray}
\label{eq5}
   \hat L\phi_{nm}(t)
   &=&L_{nm}(t)= 4 \phi_{n,m}(t)-\phi_{n+1,m}(t) -\phi_{n-1,m}(t) 
\nonumber \\
   &-&\phi_{n,m+1}(t)-\phi_{n,m-1}(t), \quad n,m=1\dots N. 
\end{eqnarray}
The local curvature is a measure for \emph{amplitude} synchronization and in the 
synchronization regime it is close to zero while in the asynchronous regime it 
is finite and fluctuating. If $g$ is the normalized probability density function 
of $|\hat L|$, then $g(|\hat L|=0)$ measures the relative size of spatially coherent 
regions in each temporal realization and characterizes the entire lattice. For a 
fully synchronized system $g(|\hat L|=0)=1$, while for a totally incoherent 
system it holds that $g(|\hat L|=0)=0$. An intermediate value of $g(|\hat L|=0)$ 
indicates the coexistence of synchronous and asynchronous SQUIDs and, therefore,
potentially interesting collective behavior. From $g$, which is time-dependent, 
we calculate the spatial extent occupied by the coherent SQUIDs which is defined 
by the integral: $g_{0}(t)=\int_{0}^{\delta}g(t,|\hat L|)d|\hat L|$, where 
$\delta=0.01L_{\mathbf{max}}$ is a threshold value distinguishing between 
coherence and incoherence and is related to the maximum local curvature 
($L_\mathbf{max}$).

In order to correspond one single value to each realization, we calculate the 
time-average $\langle g_{0}(t)\rangle$, and we plot it in the 
$(\phi_{dc},\Omega)$ parameter space. The result is shown in~Fig.~\ref{fig03}(b) 
for a coupled lattice with $\lambda=-0.025$. Yellow (bright) and purple (dark) 
regions denote a synchronized and desynchronized lattice, respectively. The cyan 
lines mark two (for visualization simplicity) of the saddle-node bifurcation 
lines of the single SQUID system of Fig.~\ref{fig03}(a). By comparing the two 
plots, it is evident that the bifurcation lines of the single SQUID almost mark 
the borders between synchronization and desynchronization of the coupled system. 
For relatively weak coupling, which is the case in Fig.~\ref{fig03}(b), this is 
plausible: When the single SQUID has one stable solution we may claim that the 
whole lattice acts like one SQUID and therefore the solution for the coupled 
system is the fully synchronized state. However, when the single SQUID loses its 
stability through the bifurcations shown in Fig.~\ref{fig03}(a), each node of 
the lattice may behave differently resulting, thus, in a desynchronized state.
For stronger coupling strengths (not shown here), the regions of incoherence around $\phi_{dc}=\pm0.5$ 
broaden, while the one around $\phi_{dc}=0$ shrinks. The structure of the parameter space (symmetry and periodicity), however, is maintained.
In terms of mean field theory, the weak coupling case essentially corresponds to the single SQUID limit, where the collective and single SQUID behavior are seen to be in close correspondence. In the other extreme of strong coupling (not addressed in this work) one may possibly introduce an order parameter and proceed in analyzing the system from the point of view of collective phenomena.


\subsection{Pattern formation}
\label{subsec:patterns}
Based on Fig.~\ref{fig03} we will study the collective 
dynamics emerging near the synchronization-desynchronization transition. 
Additionally, based on the resonance curve of the single SQUID 
(Fig.~\ref{fig02}), we select the two $\Omega$ regimes that show the most 
interesting behavior: One around the geometric resonance frequency where the 
single SQUID is extremely multistable through successive saddle-node bifurcations 
of periodic solutions, and one at lower frequencies where period-doubling takes 
place. We prepare the lattice such that the initial conditions for the 
$\phi_{nm}$ values follow a Gaussian random distribution in the interval $[-3,3]$ 
and $\dot \phi_{nm}=0.0$. As a control parameter we consider the coupling 
strength $\lambda$, which, in principle, can be tuned in an experiment by 
increasing or decreasing the distance between the SQUIDs.
\begin{figure*}[]
\includegraphics[width=\textwidth]{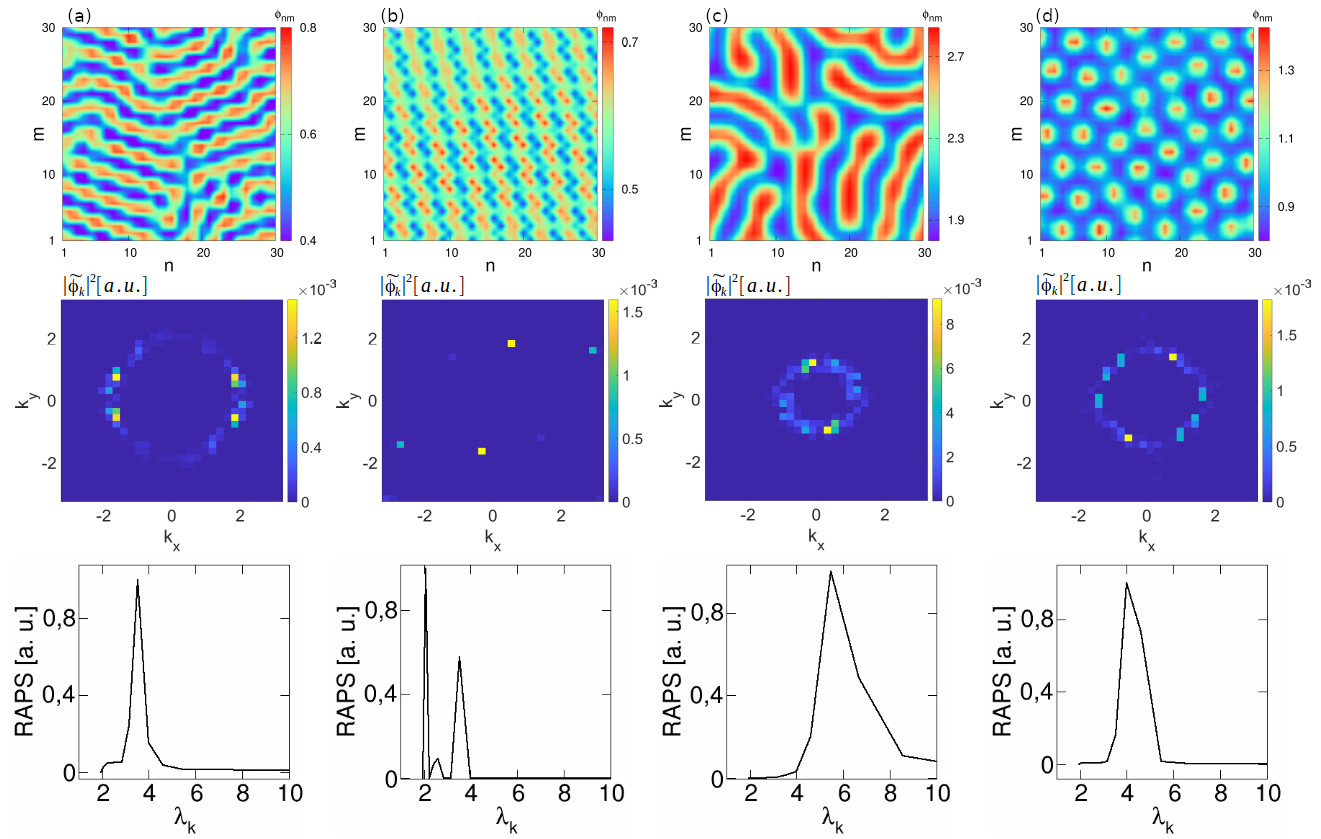}
\caption{
Top: Snapshots of the spatio-temporal patterns of the magnetic flux in a $n \times m$ SQUID lattice ($n=m=30$), corresponding to the cases \Romannum{1}-\Romannum{4} marked on Fig.~\ref{fig03}(b):  
(a) $\Omega=0.345$, $\phi_{dc}=0.5$ and $\lambda=-0.025$ (case \Romannum{1}), 
(b) $\Omega=0.345$, $\phi_{dc}=0.5$ and $\lambda=-0.039$ (case \Romannum{2}), 
(c) $\Omega=1.06$, $\phi_{dc}=0.3$ and $\lambda=-0.032$ (case \Romannum{3}), 
    and 
(d) $\Omega=1.06$, $\phi_{dc}=0.23$ and $\lambda=-0.05$ (case \Romannum{4}).  
Other parameters are: $\phi_{ac}=0.06$, $\gamma=0.024$ and $\beta=0.1369$. 
(See videos SM2(a)-(d) of the Supplementary Material for the corresponding videos).
Middle: Corresponding Fourier Power Spectra in the 2D $k$-space. 
Bottom: Radially Averaged Power Spectrum (RAPS) in $\lambda_k$.}
\label{fig04}
\end{figure*}
The results are shown in Fig.~\ref{fig04}: The top panels (a-d) show snapshots of the 
spatial distribution of the magnetic fluxes for the points 
(\Romannum{1}-\Romannum{4}) marked in the parameter space in Fig.~\ref{fig03} (b).
Figures ~\ref{fig04} (a) and (b) correspond to a low driving frequency 
value ($\Omega=0.345$) i.~e. far from the geometric resonance, where the single 
SQUID obtains low-amplitude periodic solutions and undergoes period doubling.
For a coupling strength $\lambda=-0.025$ the SQUID lattice self-organizes into a 
labyrinthine-like pattern (Fig.~\ref{fig04}(a)), while for a stronger coupling (Fig.~\ref{fig04} (b)) the collective state is a striped pattern, where smaller 
``zigzag'' patterns exist within each stripe. As mentioned previously, these patterns emerge 
from a completely random magnetic flux initialization and are, therefore, a 
result of the nonlinearity of the single SQUID and the collective dynamics of 
the coupled system.

In the middle panels, the two-dimensional Fourier power spectra 
$|\tilde{\phi_k}|^2$ are plotted in the inverse space domain. The maximum values 
of the power spectra correspond to the characteristic wavenumber 
$k=\sqrt{k^2_x+k^2_y}$ of each pattern. From $|\tilde{\phi_k}|^2$ we obtain the 
1D Radially-Averaged Power Spectrum (RAPS) \cite{Wang2007} in terms of the 
wavelength $\lambda_k=2\pi/k$, shown in the lower panels of Fig.~\ref{fig04}.
From the peaks of these RAPSs we can extract the characteristic wavelength of 
each pattern. For example, for the pattern in Fig.~\ref{fig04}(a) this value is $\simeq 3.53$, 
which is roughly the distance between two stripes in the $m$-direction of the 
corresponding plot in the top panel, in other words, the spatial period of the 
pattern. In the case of Fig.~\ref{fig04} (b), on the other hand, the RAPS obtains two maxima: 
The first one reflects the distance within the ``zigzag'' patterns inside the 
stripes ($\simeq 2$) and the second one, the distance between the stripes 
themselves ($\simeq 3.53$).

Similarly, Figs.~\ref{fig04}(c) and (d) show the patterns, and the 
corresponding Fourier power spectra in $k$-space and RAPS, obtained near the 
geometric resonance, where the single SQUID may achieve high magnetic flux 
values through saddle-node bifurcations of limit cycles. The pattern in Fig.~\ref{fig04} (c), similar to Fig.~\ref{fig04} (a), is labyrinthine-striped, but with, 
evidently, a higher characteristic wavelength $\simeq 5.45$.  For stronger 
coupling and a smaller $\phi_{dc}$ value, the emerging pattern consists of 
spots, with a characteristic wavelength equal to $4$ (Fig.~\ref{fig04} (d)).

\begin{figure*}[]
\includegraphics[width=.9\textwidth]{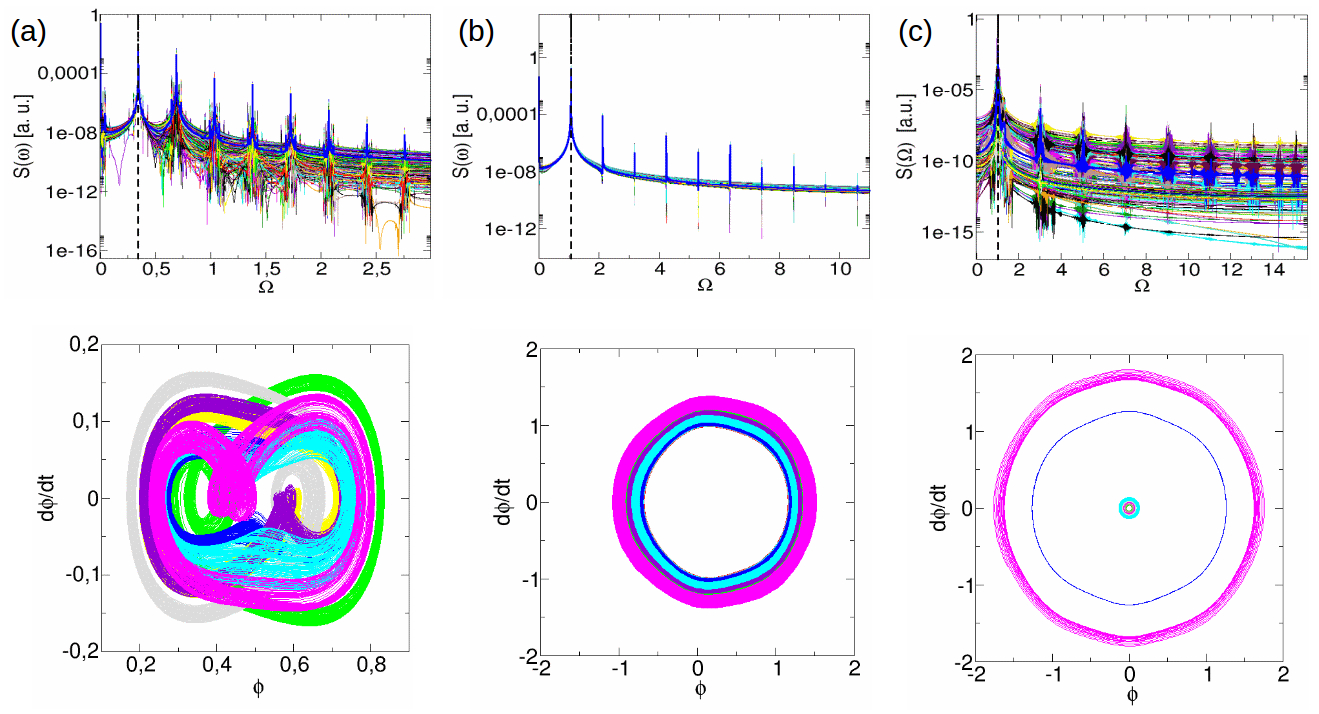}
\caption{
Top: Fourier power spectra in the frequency domain of all the SQUIDs and their
     average (thick blue line) for the spatio-temporal patterns of 
     (a) Fig.~\ref{fig04} \Romannum{1}, 
     (b) Fig.~\ref{fig04} \Romannum{1}, 
     and 
     (c) the chimera state of Fig.~\ref{fig06}(a). 
Bottom: Phase diagrams of some typical timeseries in the SQUID lattice.}
\label{fig05}
\end{figure*}

The patterns of Fig.~\ref{fig04} are spatio-temporal, and apart from a spatial 
period they also have a temporal period and corresponding frequency. These 
characteristic frequencies are given by the peaks of the Fourier power spectra 
in the inverse time domain, shown in the upper panels of Fig.~\ref{fig05}. 
Figures~\ref{fig05}(a) and (b) refer to Figs.~\ref{fig04} (a) and (c) (we have omitted the spectral analysis of cases Fig.~\ref{fig04} (b) 
and (d) because they are identical to Fig.~\ref{fig04} (a) and (c), 
respectively). We have plotted the spectra of all the SQUIDs in the lattice as 
well as their average (thick blue line). For the case of the pattern Fig.~\ref{fig04} (a), 
the spectra are very similar and the lattice is highly synchronized in frequency.
As expected, the dominant frequency is that of the driving force, marked with 
the vertical dashed line. Moreover, the spectra are rather ``noisy'' and they 
possess multiple secondary frequencies. This is typical for quasiperiodic motion 
as demonstrated by the phase diagrams in the lower panels of Fig.~\ref{fig05}(a).
The situation is similar, but ``cleaner'' for the pattern in Fig.~\ref{fig04} (c). 
As seen in the power spectra of Fig.~\ref{fig05}(b), the SQUIDs in the lattice 
are almost perfectly frequency-locked with the dominant frequency again being 
that of the driving force. The corresponding phase diagrams show, again, 
quasiperiodicity but the motion now is closer to harmonic since we are very 
close to the geometric resonance.

In correspondence with the traditionally discussed Turing patterns in Reaction-Diffusion systems, the 2D SQUID metamaterials can be characterized as partially cross-diffusive systems whose two components are the magnetic fluxes threading the loops of the SQUIDs and their time-derivatives. This can be readily inferred by taking the continuous limit of Eqs.~\ref{eq04}. Although our system exhibits similarities with classical Reaction-Diffusion systems, it also exhibits differences, with the most important being the presence of the driving force. The emergence of Turing-like patterns in forced, discrete, 2D systems such as the one considered here has not been addressed very often in the literature, see e.~g.~\cite{MUR99}. Due to the forcing term in the dynamic equations, the simple procedure to identify Turing instabilities cannot be applied. The main reason is the multistability of the individual SQUIDs, that results in a large number of periodic solutions for the SQUID metamaterial, even in the uncoupled case. Some of these solutions may be synchronized for $\lambda=0$ and may not be destroyed as the coupling is switched on. As a result, they may coexist with Turing-like patterns, indicating a complexity which cannot be handled by the simple analysis applied in classical Reaction-Diffusion systems.

\subsection{Chimera states}
\label{subsec:chimeras}
Chimeras are known to coexist with the fully synchronized state and, therefore, 
in many cases, they can be very sensitive to initial conditions. This holds for 
our system too, where chimeras can be achieved only for certain spatial 
distributions of the initial values of $\phi$. For example, in the locally 
coupled 1D SQUID array \cite{HIZ16a}, a ``sine wave'' magnetic flux distribution 
was used for the initial conditions. It was shown that the SQUIDs that were 
prepared at lower values formed the coherent clusters of the chimera state, 
while those that were initially set at higher magnetic flux values, oscillated 
incoherently. Moreover, as the ``wavelength'' of the initial magnetic flux 
distribution increased, so did the chimera state multiplicity (number of 
(in)coherent clusters). Note that in our system, since the frequency of the 
SQUID oscillators is imposed by the external driving, we are dealing with 
\emph{amplitude} chimera states~\cite{BAN18}.
\begin{figure}[ht]
\includegraphics[width=0.5\textwidth]{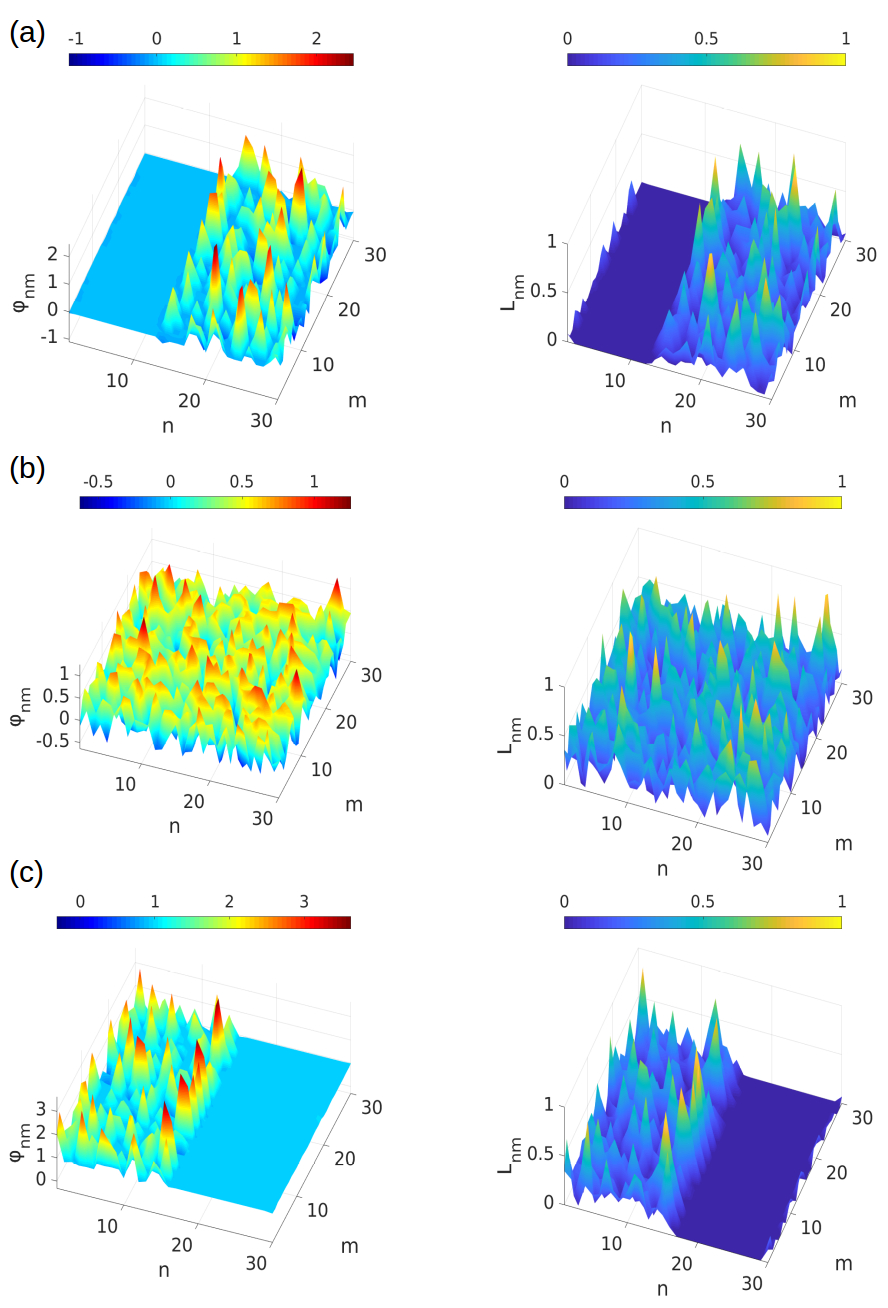}
\caption{3D snapshots of the magnetic flux (left) and the normalized discrete 
Laplacian (right) of a $30 \times 30$ SQUID lattice, for gradient initial 
conditions, $\lambda=-0.025$, $\Omega=1.007$ and different dc flux values: 
(a) $\phi_{dc}=0.0$ (see Supplementary Material for the corresponding video 
    SM2(e)), 
(b) $\phi_{dc}=0.3$, and (c) $\phi_{dc}=1.0$.  
Other parameters are: $\phi_{ac}=0.06$, $\gamma=0.024$ and $\beta=0.1369$.}
\label{fig06}
\end{figure}
Here we will employ a set of different initial conditions, inspired by 
experimental feasibility. In particular, we will use a spatial gradient for the 
magnetic fluxes $\phi_{nm}=\frac{n-1}{N-1}\phi_{\mathbf{max}}$, where 
$\phi_{\mathbf{max}}=1.5$ is the slope of the gradient, and zero values 
$(\dot \phi_{nm}=0)$ for their derivatives. Another important factor for 
achieving robust chimeras in our system is the choice of the driving frequency. 
As reported in \cite{HIZ16a}, it is crucial to be near the geometric resonance 
where the phenomenon of ``attractor crwoding''~\cite{HIZ16a} favors the emergence of such states. From section~\ref{sec1}, 
however, we know that by varying the dc flux, the snake-like form of the 
resonance curve shifts, resulting in loss of the SQUID multistability. It is 
interesting, therefore, to see what the effect of $\phi_{dc}$ will be on the 
creation of chimera states.

Figure~\ref{fig06} shows 3D snapshots of the magnetic flux (left), and their 
corresponding normalized local curvature values, when the SQUID lattice is 
prepared with gradient initial conditions. In Fig.~\ref{fig06}(a) the dc flux is 
zero and the incoherent cluster forms at the part of the lattice which is 
initially set at high magnetic flux values. On the other hand, the left half of 
the lattice is coherent and performs low-amplitude oscillations (better 
vizualized in the video SM2(e) of the Supplementary Material). By changing the 
$\phi_{dc}$ value to 0.3, the chimera state is destroyed and the collective 
state exhibits no spatio-temporal structure (Fig.~\ref{fig06}(b)). This is due 
to the fact that for this dc flux value, the single SQUID is no longer 
multistable and chimera states are not possible. By further 
increasing $\phi_{dc}$ to unity, where the single SQUID is again multistable, the chimera reappears. Interestingly, comparing to 
Fig.~\ref{fig06}(a), we observe a ``swap'' in the position of the (in)coherent 
clusters, although the initial conditions are unchanged. This is due to the fact 
that for $\phi_{dc}=1$ the ``center'' of the periodic solutions has shifted by 
$1$ and the situation is reversed compared to Fig.~\ref{fig06}(a) where 
$\phi_{dc}=0.0$. 

Apart from the single chimeras of Fig.~\ref{fig06}, we can also achieve 
multichimera states (with more than one (in)coherent clusters), simply by 
increasing the slope of the initial conditions gradient. For instance, for a 
slope of $3.5$, a multichimera state with two (in)coherent domains is formed 
(not shown here). Recently, this mechanism for the generation of chimera states 
was reported, for \emph{non-identical} coupled SQUIDs, where the gradient was in 
the dc flux distribution rather than in the initial conditions~\cite{LAZ19}. 
Such chimeras are similar to the equivalent one-dimensional structures \cite{HIZ16a},
extended in the second spatial dimension. For different special initial conditions, other types of 
chimeras are also possible, which are specific to the 2D geometry and are not present in the one-dimensional array. Here, however, we chose to focus on the ``stripe'' chimeras of Fig.~\ref{fig06}, since the gradient flux initialization is easy to achieve experimentally.

Finally, we take a look at the Fourier power spectrum of the chimera state, 
namely that of Fig.~\ref{fig06}(a), in the frequency domain. As we can see in 
Fig.~\ref{fig05}(c), the sharpest peak is located at the value of the driving 
frequency, and there are secondary broader peaks at higher frequencies too. The 
corresponding phase diagram in the panel below shows some typical solutions of 
SQUIDs in the lattice. It is clear that the coexistence of smaller and bigger 
amplitude attractors (which is absent in the patterns discussed in 
Subsection~\ref{subsec:patterns}) is the key to the emergence of chimera states 
in our system.

It should be noted that the emergence and form of chimera states does not depend crucially on the particular lattice geometry, as long as the dimensionality of the SQUID metamaterial is the same and the same initial conditions are used. For example, chimera states similar to those demonstrated in~\cite{LAZ18} for a 2D SQUID metamaterial on a Lieb lattice can be obtained for the 2D tetragonal lattice considered here, when the same initial conditions are used.

\subsection{Interplay of patterns}
As demonstrated in the previous
sections~\ref{subsec:patterns} and~\ref{subsec:chimeras}, our system is capable of exhibiting Turing-like patterns associated with the single SQUID bifurcation structure in the low coupling limit, as well as chimera states when the driving frequency is chosen close to the geometric resonance frequency. Chimeras emerge through special initial conditions and may disappear and reappear as the dc flux varies; Turing-like patterns, on the other hand,  can be obtained for a random lattice initialization and a wider parameter range. Naturally, the question arises, under which circumstances do these different patterns coexist and how do they interact with each other as the system parameters change. 

\begin{figure}[ht]
\includegraphics[width=.5\textwidth]{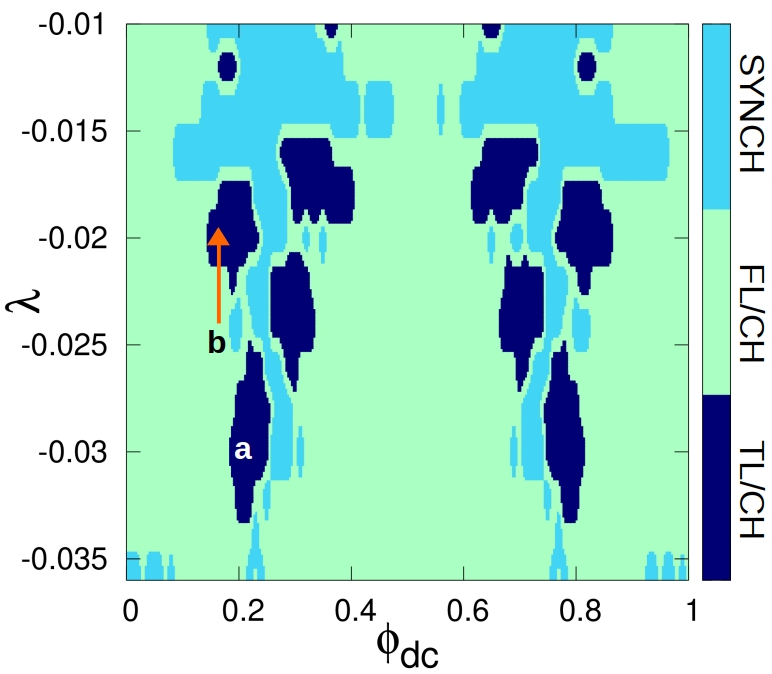}
\caption{Classification of 2D patterns in the $(\phi_{dc},\lambda)$ parameter space for $\Omega=1.03$. ``SYNCH'' stands for synchronized state, ``TL'' for Turing-like state, ``CH'' for chimera state, and ``FL'' for formless state. 
Points \textbf{a} and \textbf{b} correspond to $(\phi_{dc},\lambda)=(0.2,-0.03)$ and $(0.17,-0.025)$, respectively.
Other parameters are: $\phi_{ac}=0.06$, $\gamma=0.024$ and $\beta=0.1369$.}
\label{fig07}
\end{figure}
\begin{figure}[ht]
\includegraphics[width=.5\textwidth]{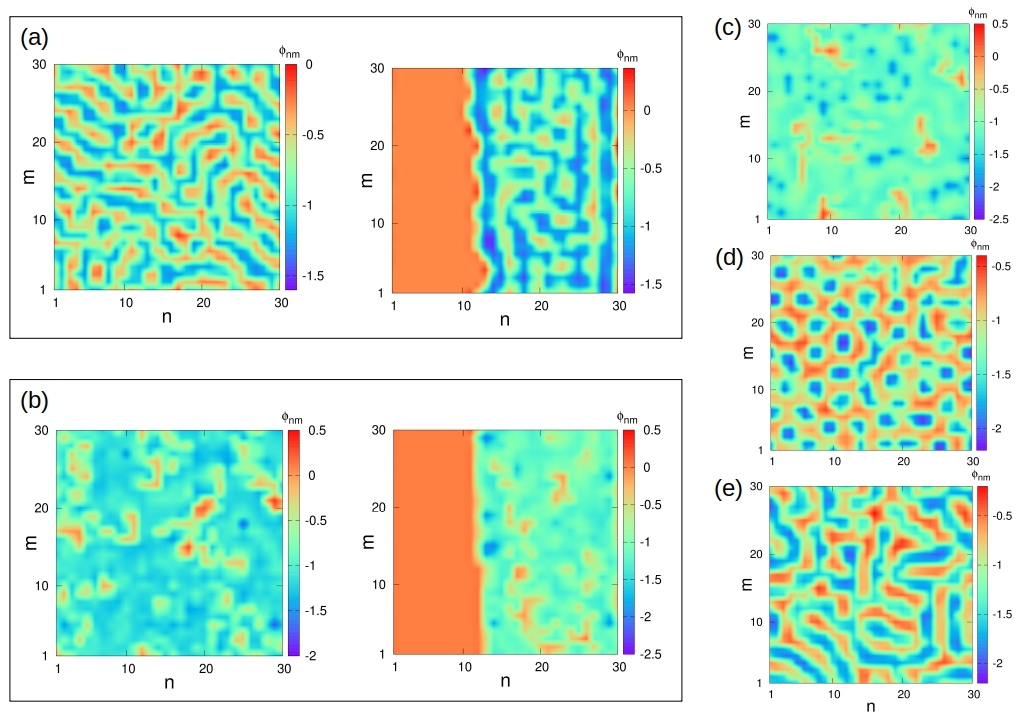}
\caption{(a) Coexisting ``TL'' state (left) and ``CH'' state (right) for point
\textbf{a} of Fig.~\ref{fig07}. (b) Coexisting `FL`` state (left) and ``CH'' state (right) for point \textbf{b} of Fig.~\ref{fig07}. Plots (c)-(d) show the evolution of the ``CH'' state of (b), as $\lambda$ increases along the arrow in Fig.~\ref{fig07}. Specifically: (c) $\lambda=-0.024$, (d) $\lambda=-0.022$, and 
(e) $\lambda=-0.02$. Other parameters are: $\Omega=1.03$, $\phi_{ac}=0.06$, $\gamma=0.024$ and $\beta=0.1369$.}
\label{fig08}
\end{figure}

Figure~\ref{fig07} shows a map of the possible patterns observed in our system, in the $(\phi_{dc},\lambda)$ parameter space, for a driving frequency close to the resonance ($\Omega=1.03$). We distinguish areas of synchronized states (SYNCH), chimera states (CH), Turing-like states (TL), and states with no clear pattern structure which we will refer to as ``formless'' states (FL). 
There are regions where different patterns may coexist, depending 
on the choice of initial conditions. We highlight two such examples, marked by points a and b. The corresponding coexisting states are shown in Fig.~\ref{fig08}(a) and (b): In Fig.~\ref{fig08}(a) a Turing-like patterns (left) coexists with a chimera state (right). Note that the unsychronized part of  the chimera has an evident spatial structure, resembling a ``half'' Turing-like pattern. On the other hand, in Fig.~\ref{fig08}(b), the chimera state (right) has a rather formless desynchronized domain and coexists with an unstructured pattern shown in the left. 

These coexisitng patterns are robust and do not transform between each other, as our long simulations (of the order of $10^4$ periods) can confirm.
The transformation between states, however, can be achieved by varying the system parameters, namely the coupling strength and the dc flux, as shown in Fig.~\ref{fig07}. Specifically, a chimera state may evolve into a Turing-like pattern, but not the reverse since chimeras require special initial conditions in order to occur. Additionally, a formless state may also change into a Turing-like pattern and vice versa. By performing a continuation of states while varying the coupling strength $\lambda$ in the direction of the arrow in Fig.~\ref{fig07}, we can see that the chimera state of Fig.~\ref{fig08}(b) loses its structure and becomes a formless state in Fig.~\ref{fig08} (c). In turn, this state, by further increase of $\lambda$, evolves into a Turing-like pattern as depicted in Figs.~\ref{fig08} (d) and (e). Interestingly, the coexistence of chimeras and Turing-like patterns has been reported before in two-dimensional networks of nonlocally coupled neurons in the low coupling limit~\cite{SCH17} but, in general, is a question yet to be explored.

\section{Conclusions}
In conclusion, we have shown that a 2D SQUID lattice with nearest neighbor 
interactions is capable of exhibiting a rich menagerie of Turing-like pattern forming states.
In the low coupling limit, this collective behavior emerges near the transition from synchronization to 
desynchronization where the single SQUID undergoes complex bifurcations. 
Moreover, near the geometric resonance, we observe 2D chimera states, as a 
result of the extreme multistability of the single SQUID. What is interesting is 
that by proper choice of initial conditions and tuning of the dc flux of the 
driving force, we are able to control the multiplicity and position of the 
chimera states, respectively. Additionally, in certain regions of the parameter space, chimeras may coexist with Turing-like patterns and also evolve into such,
by proper tuning of the relevant parameters. Recent experiments~\cite{ZHU19} on the imaging of 
collective states in SQUID metamaterials through laser scanning microscopy (LSM 
technique), are very promising in terms of verifying our theoretical findings in 
the lab.


\section{Supplementary Material}
See Supplementary Material for the videos corresponding to Fig.~\ref{fig02}, the patterns of 
Fig.~\ref{fig04}, and the chimera state of Fig.~\ref{fig06}(a).

\section*{Acknowledgement}
This work was  financially supported by the Ministry of Education and
Science of the Russian Federation in the framework of Increase
Competitiveness Program of NUST ``MISiS'' (grant No. K3-2018-027).
JH and NL acknowledge support by
the General Secretariat for Research and Technology (GSRT)
and 
the Hellenic Foundation for Research and Innovation (HFRI) (Code: 203).
JH would also like to thank Jan Sieber for helping with the continuation tool.


\end{document}